\documentclass[12pt]{iopart}
\input{psfig.sty}

\def\apj{{ApJ}}

\def\mnras{{MNRAS}}

\newlength{\tskip}
\newlength{\colwidth}

\newcommand{\beq}{\begin{equation}}
\newcommand{\eeq}{\end{equation}}
\newcommand{\beqa}{\begin{eqnarray}}
\newcommand{\eeqa}{\end{eqnarray}}

\newcommand{\bn}{\hat{\bf n}}

\newcommand{\rad}{r}    

\newcommand{\isw}{{\rm ISW}}

\newcommand{\len}{{\rm len}}

\begin{document}
\title[Did WMAP see Moving Local Structures?]{Did WMAP see Moving Local Structures?}
\author{ Asantha Cooray, Naoki Seto}
\address{Department of Physics and Astronomy, University of California, Irvine, CA 92617}

\begin{abstract}
The divergence of the momentum density field of the large scale structure 
generates a secondary anisotropy contribution to the Cosmic Microwave Background (CMB).
While the effect is best described as a non-linear extension to the well-known integrated Sachs-Wolfe
effect, due to mathematical coincidences, the anisotropy contribution is also described as the
lensing of the dipole seen in the rest-frame of a moving mass.
Given the closeness, there is a remote possibility that local concentrations of mass in the form of
the Great Attractor and the Shapley concentration generate large angular scale fluctuations in CMB
and could potentially be responsible, at least partly, for some of the low-multipole anomalies in WMAP data.
While the local anisotropy contribution peaks at low multipoles, for reasonable models of 
the mass and velocity distributions associated with local super structures we find that the amplitude of temperature anisotropies
is at most at a level of 10$^{-2}$ $\mu$K and is substantially smaller than primordial fluctuations.
It is extremely unlikely that the momentum density of local mass concentrations is responsible for any of the
large angular scale anomalies in WMAP data. 
\end{abstract}

\maketitle

\section{Introduction}
The presence of several large-scale, or ``low-$\ell$'', 
anomalies in cosmic microwave background (CMB) 
data from  Wilkinson Microwave Anisotropy Probe (WMAP; Bennett et al. 2003) is now routinely discussed (see, Copi et al. 2005 for a 
comprehensive study). Anomalies such as the lack of power  at angular scales
above $\sim$ 60$^\circ$ (Spergel et al. 2003) are consistent with what were previously seen with
Cosmic Background Explorer's Differential Microwave Radiometer (COBE-DMR; Hinshaw et al. 1996) data.
The statistical significance of anomalies such as the alignment of the quadrupole and the 
octupole (de Oliveira-Costa al. 2004; Schwarz et al. 2004) is still uncertain due to deficiencies in our understanding of foreground 
contamination (e.g., Slosar \& Seljak 2004; Bielewicz et al. 2005).

Suggestions for the lack of large-scale power include finite universe models with non-standard topology (e.g., Luminet et al. 2003),
a cut-off in the initial perturbations at scales corresponding to the present-day horizon and above (e.g., Efstathiou 2004),
and perturbations in dark energy (e.g., Gordon \& Hu 2004). 
As the existing measurements are cosmic variance limited, further CMB observations are unlikely to
improve the statistical significance of low power detection at large scales. The statistics may slightly be improved
with a cosmic shear analysis of future all-sky high resolution CMB maps (Kesden et al. 2003).
Suggestions for the alignment between the quadrupole and the octupole
include a local source of anisotropy contribution to CMB, 
such as the Sunyaev-Zel'dovich effect from the local supercluster intergalactic medium (Abramo \& Sodre 2003; see, Hansen et al. 2005) or
gravitational lensing of the CMB dipole (Vale 2005), and  a modification to large angular scale CMB anisotropies through
anisotropic fluctuations in the dark energy component (Gordon et al. 2005).

The late-time integrated Sachs-Wolfe (ISW; Sachs \& Wolfe 1967) effect associated with time evolving 
potentials, due to gravitational growth of the large scale structure density field, 
is now well known. This effect generally contributes at large angular scales and is associated with the difference in
photon frequency, or energy, as photons fall in and climb out of potential wells of the large scale mass distribution.
In the limit that the potentials are fixed in time or structures become non-linear and virialize, 
the blue shift experienced by photons falling in to gravitational potentials exactly cancels out with the redshift experienced by photons 
when climbing out of the potentials and there is no contribution to CMB anisotropies. 
On the other hand, the motion of these mass concentrations
generates a higher order contribution to CMB anisotropies (Rees-Sciama effect; Rees \& Sciama 1968).
This non-linear correction to ISW effect can be  described as
the divergence of the momentum density field associated with large-scale mass distribution (Seljak 1996; Cooray 2002; see simulations in
Tulie \& Laguna 1995). Incidently, this non-linear ISW contribution to CMB temperature anisotropies is essentially same
as those produced by ``moving'' gravitational lenses (Birkinshaw \& Gull 1983; Gurvits \&  Mitrofanov 1986).
A simple description based on lensing, however, is not appropriate as the effect 
does not involve a lensing of a ``true'' source as the source only exists in the rest-frame of a moving mass. The lensing description also becomes less useful
in the case of significant
internal motions rather than a coherent bulk velocity
when an overdensity is collapsing to form a virialized structure.
A fluid dynamical approach to calculate the non-linear ISW effect
for such a scenario is described in Lasenby et al. (1999; with numerical
calculations in Dabrowski et al. 1999). The  cosmic string analogue of this non-linear effect is described in Kaiser \& Stebbins (1984).

Due to closeness, local mass concentrations project on the sky at large angular scales and there is a remote possibility
that the momentum density field of the local group of galaxies is responsible for certain low-multipole anomalies in WMAP data.
To answer this possibility, we calculate the expected contribution to
CMB anisotropies by representative local super structure based on
published models of mass and velocity distributions out to $\sim$ 30,000 km sec$^{-1}$ in the literature.
These are summarized in the next section.
We concentrate on overdensities such as the Great Attractor and the Shapley concentration of galaxies, and
generally find that the anisotropy contribution is at least two to three orders of magnitude lower than what is required to
explain the quadrupole and the octupole alignment. As discussed in \S~3, we conclude that the local structure is unlikely to be
responsible for the WMAP anomalies.

\begin{figure*}[!t]
\centerline{\psfig{file=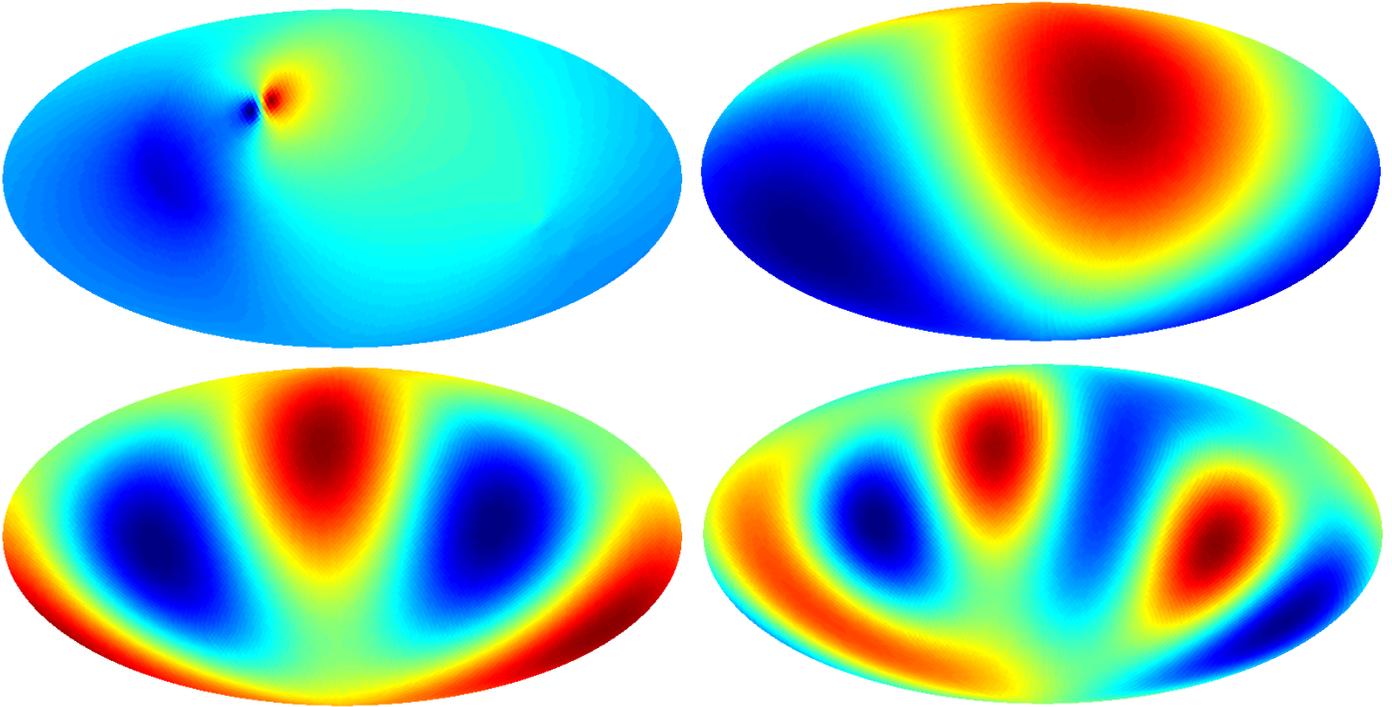,width=7.3in,angle=0}}
\caption{The local universe contributions to CMB anisotropies. We consider two mass concentrations:
the Great Attractor (GA) at a distance of
32 h$^{-1}$ Mpc and the Shapley concentration at a distance of
145 h$^{-1}$ Mpc.  In the top-left panel, we show the temperature anisotropy maps
produced by the motion of these two mass clumps; the large-angle dipole pattern is 
from the GA, while the smaller concentrated dipole pattern is associated with the 
SC. The negative anisotropy (a redshift corresponding to a
temperature decrement shown in blue) is in the same direction as the velocity.
The dipoles are aligned towards the same direction as we assume the bulk flow within 140 Mpc
to model their motions and, furthermore, the two dipoles are located in the same area as 
centers of both mass concentrations are in the same general area on the sky (see text for details).
The maximal temperature fluctuation is of order $\sim$ 0.5 $\mu$K.
In the top-right, bottom-left, and bottom-right panels,
we show the resulting patterns related to the dipole, quadrupole and the octupole.
The power spectrum of anisotropies generated by the motion of these structures is shown in Fig.~3.}
\label{fig:wmap}
\end{figure*}

\section{Local Universe Contributions}

The integrated Sachs-Wolfe effect (Sachs \& Wolfe 1967) results from
the late time decay of gravitational potential fluctuations. The
resulting
temperature fluctuations in the CMB can be written as
\begin{equation}
\frac{\Delta T^\isw(\bn)}{T} = -2 \int_0^{\rad_0} d\rad \dot{\Phi}(\rad,\bn \rad) \, ,
\end{equation}
where the overdot represents the derivative with respect to conformal
distance (or equivalently look-back time).

For non-linear structures, one can employ the continuity equation
and the Poisson equation. This allows one to relate the time derivative of the potential field
to divergence of the momentum field (for details, see Seljak 1996; Cooray 2002).
The resulting temperature anisotropy term  includes both linear and  non-linear  contributions:
$\Delta T/T \propto [\dot{a}/{a}\delta + \nabla \cdot (1+\delta) {\bf v}]$.
The linear ISW effect comes as a contribution of two terms, $(\dot{a}/{a}\delta +\nabla \cdot {\bf v})$,
while the non-linear contribution involves the second-order term that appears as
a product of the overdensity of the mass fluctuation and it's velocity:
$\Delta T^{\rm nl} \propto \nabla \cdot \delta {\bf v}$. 
Rewriting the density field in terms of potential fluctuations using 
the Poisson equation, putting all terms exactly (see, e.g., Cooray 2002) 
we can write the anisotropy contribution as:
\begin{eqnarray}
\frac{\Delta T^{\rm nl}}{T} &=& -2 \nabla \cdot (\Phi {\bf v}) \nonumber \\
&\approx&-{\bf v_\perp} \cdot  \left(2\nabla_r \Phi\right) \nonumber \\
&=&-v \sin \alpha \delta_\len \cos \phi  \, .
\label{eqn:temp}
\end{eqnarray}
The approximation in the second line above assumes that potential fluctuations are embedded in a
velocity field with much larger coherence scale so that gradients
in the velocity field do not contribute to temperature
anisotropies. This is the case where a virialized mass, such as a galaxy cluster, is in motion as a single object.
Massive structures that have not virialized will involve infall velocities between different subgroups
of the same region. With respect to collapsing clusters, a
fluid dynamical approach to calculating the background anisotropy is
discussed in Lasenby et al. (1999; Dabrowski et al. 1999).
Since temperature anisotropy contribution is determined by
the gradient of the momentum density field, even in a case where infall velocities are
large, as the subclumps are likely to have small masses, the resulting effect on CMB may not be significant. In addition
to the ``bulk flow'' effect, we also considered  internal motions and we will
comment on how the signal changes in the presence of internal velocities later.

As written above, the description in terms of a ``moving lens'' becomes clear in the limit
that the whole mass is moving with a single velocity. In this case, the angular gradient applies only to the
potential which is what is encountered in gravitational lensing studies as the deflection angle:
$\delta_\len = 2\nabla_r \Phi$. For single mass concentrations, the deflection angle can be written as $\delta_\len=4 GM/c^2R$
where $R$ is the typical size scale of the mass. Note that there is
no contribution to temperature anisotropy  from the gradient of the potential along the line of sight. 
Similarly, the contributing component of the velocity field is the one on the sky, i.e., the transverse velocity.
In above, $\alpha$ is the angle between the line of sight and the
velocity field and $\phi$ is the position angle from the observer.
This description involving a moving lens is not strictly correct.
The effect does not involve lensing of a ``true'' background source by the foreground moving
object. The source only exists in the rest-frame of the moving mass in the form of a dipole 
anisotropy with the amplitude $\Delta T/T = v \sin\alpha/c$ (Birkinshaw \& Gull 1983; Gurvits \& Mitrofanov 1986). 
The lack of a true source also creates a problem when considering the typical mapping between the source and images in lensing, since such mapping
depends on the ratio of distances between the lens and the source, and between the observer and the source. On the other hand, the anisotropy amplitude of the non-linear ISW effect is  independent of the exact location of the ``moving lens'' as in the case that the background source  is at an infinite distance away. The location only determines the angular scale associated with temperature fluctuations.
While a lensing description is routinely employed in the literature in calculating this effect
(e.g., Aghanim et al. 1998; Molnar \& Birkinshaw 2000), no photons are {\it deflected} by the moving mass. 
The temperature anisotropy pattern for  a given moving mass is a dipole 
centered on the object on a direction determined simply by the transverse velocity component of the mass.

\begin{figure*}[!t]
\centerline{\psfig{file=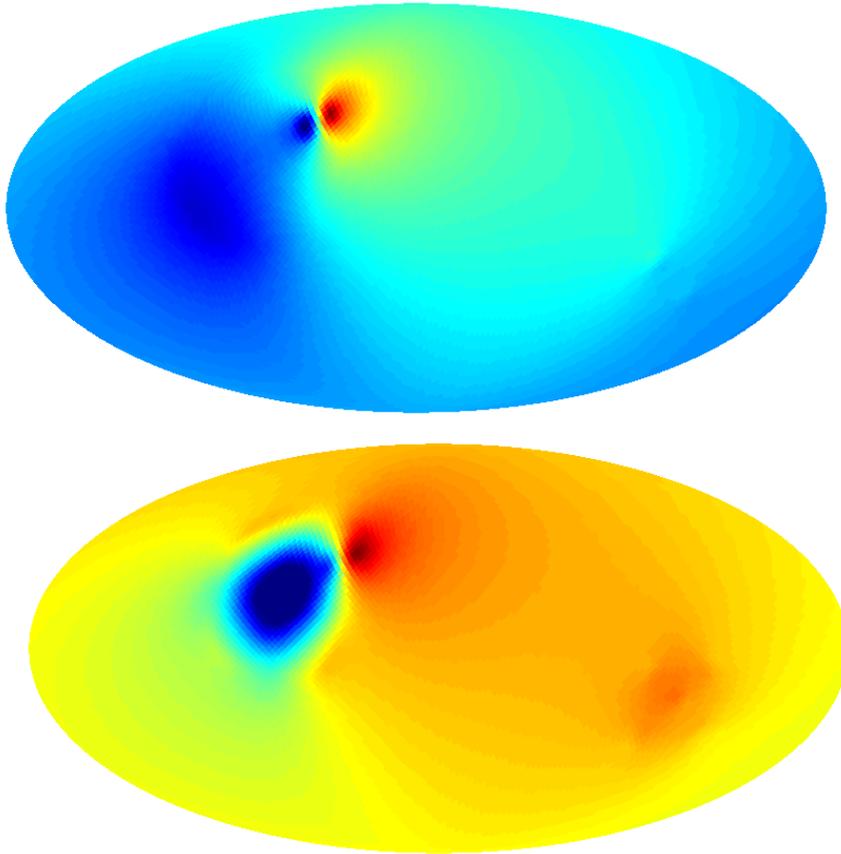,width=4.5in,angle=0}}
\caption{The correction  resulting from internal motions within the GA.
The top map shows the bulk motion contribution (same as Figure~1), while
in the bottom map, we show the anisotropy pattern that includes a model
description for internal motions within the GA, in addition to bulk
motion contribution from the GA and the SC. Internal motions
increase the over amplitude of temperature anisotropies and
the fluctuation power at large angular scales as shown in
Fig.~3. Infalling velocities generally lead to temperature decrement (shown in blue) towards the cluster center
and corresponds to an additional redshift to the CMB photons.}
\label{fig:internal}
\end{figure*}

To calculate the local contribution to CMB anisotropies through motions of mass concentrations 
following equation~(\ref{eqn:temp}), we make use of observationally derived estimates on the mass
and their velocities. We consider three overdensities within $\sim$ 30,000 km s$^{-1}$
named Great Attractor (GA; at 3200 $\pm$ 260 km s$^{-1}$; Tonry et al. 2000),
Shapley concentration (SC; at 4200 km sec$^{-1}$; Hudson et al. 2004), and S300 (Bardelli et al. 2000).
We locate the GA at $l=289^\circ$,  $b=19^\circ$ and at a distance of $32h^{-1}$Mpc, and fix its
mass at $8\times 10^{15}M_\odot$ (Tonry et al. 2000) with a size 
$14.0h^{-1}$Mpc (Hudson et al. 2004).  We center the SC on rich cluster
cluster   Abell 3358 at $l=312^\circ$, $b=32^\circ$ at a distance of $145h^{-1}$Mpc
with a mass of $6.6\times  10^{15}h^{-1}M_\odot$ and a size of $14.1h^{-1}$Mpc (Bardelli et al. 2000).

To describe velocities of these two systems, we use the mean bulk
flow of our local universe within $120h^{-1}$Mpc estimated from the SMAC
(Streaming Motions of Abell Clusters) sample (Hudson et al. 2004). Its
direction is $l=273^\circ$ and  $b=6^\circ$ with a magnitude of 372km sec$^{-1}$. With
these configurations,  projection factors related to the bulk flow velocity of
these two  masses  are $\sin \alpha=0.35$ and 0.70 for GA and SC, respectively.
We also considered the overdensity S300 of Bardelli et al. (2000) at $\sim$ 300h$^{-1}$ Mpc, but due to uncertain
parameters did not include it explicitly. As we describe below, our calculation related to
large angular scale anisotropies is unaffected whether this mass concentration
is considered or not. 

Both the GA and the SC are not virialized objects, and we adopt a simple top-hat
density profile to describe them. As we are primarily interested in the large angular
(smaller $l$) scale of the temperature map, details related 
to internal density profile are
not important here. The pattern $\Delta T/T$  of each
structure in equation~(2) has a dipole profile around  the center of the mass.  The
characteristic angular scale $\Delta \theta$ is the spatial size of the mass concentration divided by its
distance. Due to the cancellation of positive and negative temperature fluctuation regions of the dipole pattern,
a small-scale dipole pattern is not expected to
produce large  angular scale anisotropies at multipoles $\ell \ll (\Delta \theta)^{-1}$. 
This is the reason why a more distant structure like S300 (Bardelli et al. 2000)
is not expected to change our result qualitatively at low $\ell$, 

It is well known that infalling structures could also produce
significant temperature anisotropy (Saez et al. 1993; Meszaros \&
 Molnar 1996; Dabrowski et al. 2000, and references therein), as can be seen from
 eq.~(1) easily.  To capture some aspects of internal motions, but still using same calculational method as the one involving
 the momentum field, we assume that the GA is forming by infalling subclumps,
 and apply eq.~(2) to each one of them. As the pattern produced by eq.~(2) is linear with
 respect  to the   momentum, we superposed the infalling effect of each subclump on the
 pattern induced by the overall bulk velocity of GA mentioned above.  
 We put four subclumps (masses $2\times 10^{15}M_\odot$) on the plane
 perpendicular to the direction of GA center, and took their velocities towards
 the center of GA to be at 1000 km sec$^{-1}$ (Tonry et al. 2000) and their distances away from the center 
at $7h^{-1}$ Mpc. 

\section{Results}

In Figure~1, we show the anisotropy pattern produced by the GA and the SC, as well as patterns of
dipole, quadrupole and the octupole. The maximal effect is
\begin{equation}
{\Delta T \over T} = 0.5\;  \mu{\rm K} \; \left(\frac{v_\perp}{300\; {\rm km\; sec^{-1}}}\right)
\left(\frac{M}{10^{16}\; M_{\rm sun}}\right)\left(\frac{R}{10\; {\rm Mpc}}\right)^{-1} \, .
\end{equation}
With the GA and the SC combined, the multipole moments, centered on the Galactic center,  have amplitudes of
 $a_{20} = 4.2 \times 10^{-2} \mu$K, $|a_{21}|=3.6 \times 10^{-2} \mu$K,
$|a_{22}|=5.8 \times 10^{-2} \mu$K in the case of the quadrupole, while the octupole moments are
generally a factor of 3 lower. Using the temperature fluctuation map shown in the top-left panel of Fig.~1
we also calculated the angular power spectrum of temperature anisotropies. These are summarized in Fig.~3,
where we also show the angular power spectrum of temperature fluctuations measured by WMAP (Spergel et al. 2003). In Fig.~2, we show the correction
related to internal motions within the GA. As shown in Fig.~3, 
internal motions lead to an
increase in the temperature fluctuation power at large angular scales.
Our estimates are fully consistent with previous calculations in the literature, especially in the context of the local group contribution to the
quadrupole anisotropy measured by COBE (Saez et al. 1993;
Meszaros \& Molnar 1996, and references therein).

The resulting contributions from the moving local 
mass concentrations are lower by two to three orders
of magnitude relative to primordial fluctuations. 
Including the internal motions with the GA, for example, lead to a slight (a factor of a few) increase in the
overall signal, but still below a level to be of any significance. Note that the local group motions
have been considered in  the past to explain, for example, COBE quadrupole (Saez et al. 1993),
and the estimates we make here are consistent with those previous estimates.
The WMAP quadrupole and octupole moments are
at the level of $\sim 10$ $\mu$K. It is clear that local mass concentrations, due to their motions, do not
generate an adequate anisotropy to explain the low-$\ell$ anomalies in WMAP data.

Since the momentum density is the quantity that determines the anisotropy,
if moving local mass clumps are the reason, then the product $Mv$ should generally be higher by a factor of
at least $10^3$. While the mass and velocity estimates used here are from the literature, an increase in either mass or velocity,
or both, by such a factor is likely to be disfavored significantly by local group data.
Such a high mass would also run in to other problems, as it would contain essentially the same, or mass, expected within the
Hubble volume, or the velocities will become close to relativistic values. 
In addition to secondary effects based on modifications to gravity, scattering of photons also
lead to modifications and among these effects Sunyaev-Zel'dovich (SZ) effect is the dominant contribution.
The SZ effect from the local group has been suggested as a possibility for the large angular scale
anomalies (Abramo \& Sodre 2003), but improved models indicate that the local group 
is unlikely to produce a sufficiently large SZ effect  to modify the quadrupole and the octupole.
In combination, both SZ and moving  mass  overdensities  of the local group, either bulk flow or motions within,
are unlikely to be responsible for any of the anomalies seen in WMAP data. 

\begin{figure}[!t]
\centerline{\psfig{file=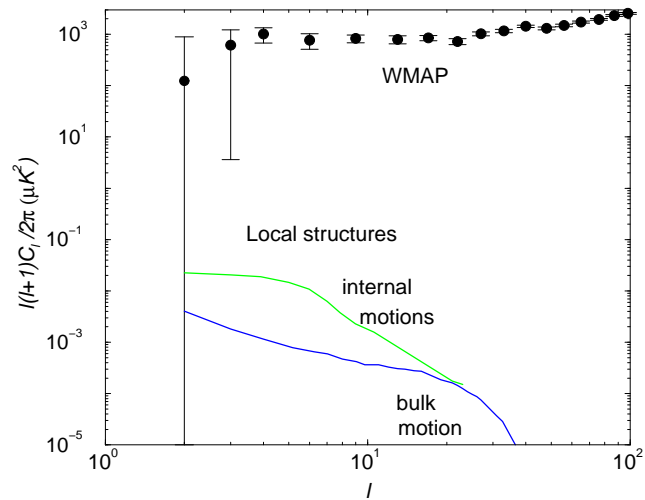,width=3.3in,angle=-90}}
\caption{Angular power spectrum of CMB anisotropies produced by local mass concentrations. This power spectrum is measured directly
on temperature fluctuation map shown in Fig.~1 (top-left panel). The curve labeled internal motions is the increase in signal
expected based on a simple estimate of motions within the GA (Fig.~2), 
in addition to the signal generated by the bulk-flow. 
For comparison, we also show the power spectrum of
CMB anisotropies measured in WMAP data (Spergel et al. 2003). The temperature fluctuation power spectrum 
of moving local mass clumps are four to five orders of magnitude smaller than primordial anisotropies.}
\label{fig:cl}
\end{figure}

\section{Conclusions}

The divergence of the momentum density field of the large scale structure 
generates a secondary anisotropy contribution to the Cosmic Microwave Background (CMB).
While the effect is best described as a non-linear extension to the well-known integrated Sachs-Wolfe
effect, due to mathematical coincidences, the anisotropy contribution is also described as the
lensing of the dipole seen in the rest-frame of a moving mass.
Given the closeness, there is a remote possibility that local concentrations of mass in the form of
the Great Attractor and the Shapley concentration generate large angular scale fluctuations in CMB
and could potentially be responsible, at least partly, for some of the low-multipole anomalies in WMAP data.
While the local anisotropy contribution peaks at low multipoles, for reasonable models of 
the mass and velocity distributions associated with local super structures we find that the amplitude of temperature anisotropies
is at most at a level of 10$^{-2}$ $\mu$K and is substantially smaller than primordial fluctuations.
It is extremely unlikely that the momentum density of local mass concentrations is responsible for any of the
large angular scale anomalies in WMAP data. 

\section{Acknowledgments}
A.~C.\ thanks Dragan Huterer for a routine to plot anisotropy maps and Chris Vale for a useful discussion
related to the dipole lensing effect. We acknowledge the use of publicly available Healpix package
(G\'orski et al. 2005).

\end{document}